\documentclass{appbav}
\usepackage{epsfig}
\usepackage{cite}

\def\titleheading{\small\rm DCPT/05/\dcptNo, IPPP/05/\ipppNo \hfil 
 hep-ph/\hepNo}

\def\dcptNo{140}
\def\ipppNo{70}
\def\hepNo{0511112}

\def\dcptNo{138}
\def\ipppNo{69}
\def\hepNo{0511112}

\newcommand{\beq}{\begin{equation}}
\newcommand{\eeq}{\end{equation}}
\newcommand{\bea}{\begin{eqnarray}}
\newcommand{\eea}{\end{eqnarray}}
\newcommand{\nn}{\nonumber}
\newcommand{\gsim}{\raisebox{-0.07cm}{$\:\:\stackrel{>}{{\scriptstyle
 \sim}}\:\: $} }

\newcommand\Nf{n_{\rm f}^{}}
\newcommand\qV{\mbox{\boldmath $q$}}
\newcommand\PV{\mbox{\boldmath $P$}}

\newcommand{\MSb}{$\overline{\mbox{MS}}$}
\newcommand{\DISg}{$\mbox{DIS}_\gamma$}

\def\a{\alpha}   \def\d{\delta}
\def\e{\epsilon}  \def\g{\gamma} 
    
   \def\x{\xi}

\begin{document}
\title{\sc PHOTON-PARTON SPLITTING FUNCTIONS AT$_{\, _{\! _{\!}}}$\\
THE NEXT-TO-NEXT-TO-LEADING ORDER OF QCD%
\thanks{Presented by A.V. at {\sc Photon 2005}, Warsaw, 
August/September~2005}%
}
\author{A. Vogt
\address{IPPP, Department of Physics, University of Durham\\
South Road, Durham DH1 3LE, United Kingdom\\[4mm]}
S. Moch
\address{Deutsches Elektronensynchrotron DESY\\
Platanenallee 6, D--15735 Zeuthen, Germany\\[4mm]}
J. Vermaseren
\address{NIKHEF Theory Group\\
Kruislaan 409, 1098 SJ Amsterdam, The Netherlands}
}
\maketitle
\begin{abstract}
\vspace*{-4mm}
\noindent
We have calculated the splitting functions governing the evolution of the 
unpolarized parton distributions of the photon at the next-to-next-to-leading
order (NNLO) of massless perturbative QCD. The results, presented here mainly 
in terms of compact and accurate parametrizations, are consistent with our 
previous approximations based on the lowest six even-integer Mellin moments. 
Consequently the NNLO corrections are small in both the \MSb\ and the \DISg\ 
factorization schemes at momentum fractions $x \gsim 0.1$.
\end{abstract}

\vspace*{8mm}
\noindent
The partonic structure of the photon, accessed in particular by the deep-%
inelastic structure function $F_2^{\,\g}$, is a classic subject in perturbative
QCD. The leading-order (LO) and next-to-leading order (NLO) expressions for the
photon-parton splitting functions $P_{\rm p\g}$, p = q, g, and the coefficient
functions $c_{\:\! 2,\g}^{}$ have been known for a long time 
\cite{Witt,BaBu,WJSWW,GR83,FP92,GRVg1}. 
A couple of years ago we have presented \cite{Moch:2001im} the corresponding 
next-to-next-to-leading order (NNLO) corrections, albeit with one important
qualification concerning the ${\cal O}(\a \a_{\rm s}^2)$ splitting functions:
at that time we were only able to derive the lowest six even-integer Mellin
moments, $N = 2,\ldots,12$, which are sufficient for a reliable reconstruction
only at momentum fractions $x \gsim 0.05$. 

Using methods and results derived in the meantime for the hadronic case 
\cite{Moch:2004pa,Vogt:2004mw,Vermaseren:2005qc}, we are now able
to present the complete results for the splitting functions and coefficient
functions to order $\a \a_{\rm s}^2$, thus finalizing the NNLO description and
providing an important partial result for $F_2^{\,\g}$ at N$^3$LO.
In the present brief contribution, we confine ourselves to the NNLO splitting 
functions, presenting accurate compact parametrizations of those results which 
are rather lengthy. Furthermore we take a first look at their numerical 
effects.  A more detailed account, including the exact results and the 
$\a \a_{\rm s}^2$ photonic 
coefficient functions for $F_2$ and $F_L$, will be presented elsewhere
\cite{MVVprep}. 

\vspace*{1mm}
At lowest order in the electromagnetic coupling $a_{\rm em} \equiv \a_{\rm em}
/(4\pi)$, the parton distributions of the photon are subject to the evolution 
equations of the form
\beq
\label{qevol}
  \frac{d\, \qV^\g}{d \ln \mu^{\:\! 2}} \: = \:
  \PV^\g + \PV \otimes \qV^\g \:\: ,
\eeq
where $\mu$ represents the \MSb\ factorization and renormalization 
scale, and $\otimes$ stands for the Mellin convolution in the momentum 
variable. For brevity writing out only the flavour-singlet case, $\qV^{\g}$ is 
given by
\beq
\label{qs}
  \qV^\g \: = \: \left( \begin{array}{c}
              \!q_{\:\!\rm s}^{\,\g}\! \\ \!g^{\,\g}\! \end{array} \right)
  \:\: , \quad\quad
  q_{\:\!\rm s}^{\,\g} \:\equiv\:\sum_{j=1}^{\Nf}\, 
                 ( q_j^{\,\g}+\bar{q}_j^{\,\g})
         \: = \: 2\,\sum_{j=1}^{\Nf} q_j^{\,\g} \:\: .
\eeq
$\Nf$ is the number of active flavours, and the splitting-function matrices are
\beq
\label{Pmat}
  \PV^\g \: = \: \left( \begin{array}{c}
    P_{\rm q\g} \\ P_{\rm g\g} \end{array} \right)
  \:\: , \quad\quad
  \PV \: = \: \left( \begin{array}{cc}
    P_{\rm qq} & P_{\rm qg} \\
    P_{\rm gq} & P_{\rm gg} \end{array} \right) \:\: .
\eeq
The expansions of the photon-parton and parton-parton splitting
functions up to NNLO read, with $a_{\rm s} \equiv \a_{\rm s}/(4\pi)$,
\bea
\label{Pgexp}
  \PV^\g & = & 
      a_{\rm em}\, \left( \PV_\g^{(0)} + a_{\rm s}\,
      \PV_\g^{(1)} + a_{\rm s}^2 \, \PV_\g^{(2)} \right)
      \\
  \PV \:\: & = & \quad\:
      a_{\rm s} \, \PV^{(0)} \,
    + a_{\rm s}^2\, \PV^{(1)} + a_{\rm s}^3 \, \PV^{(2)} \:\: .
\label{Phexp}
\eea
 
The hadronic quantities $\PV^{(2)}$ in Eq.~(\ref{Phexp}) can be found in 
Refs.~\cite{Moch:2004pa,Vogt:2004mw}. Our new results for the photonic 
non-singlet (see Ref.~\cite{Moch:2001im} for notational details) and gluon 
\MSb\ splitting functions $\PV_\g^{(2)}$ in Eq.~(\ref{Pgexp}) can be 
parametrized as
\bea
  \lefteqn{
  \d_{\rm ns}^{-1} P^{(2)}_{\rm ns \g}(x)\! \:\: \cong \:\:
       128/27\: L_1^4 + 112/9\: L_1^3 + 175.3\, L_1^2 + 142.3\, L_1
     + 1353 } 
  \nn \\[-0.2mm] & & \; \mbox{}
     - 1262\, x + 449.2\, x^2 - 1445\, x^3
     - L_0 L_1\, ( 162.7\, L_0 + 195.4\, L_1 )
     + 1169\, x L_0 
  \nn \\ & & \; \mbox{}
     + 50.08\, (1-x) L_1^3
     + 744.6\, L_0 + 201.6\, L_0^2 + 80/3\: L_0^3 + 64/27\: L_0^4
  \nn \\ & & \!\! \mbox{} + \Nf \: \big\{
     - 32/27\: L_1^3 - 11.858\, L_1^2 - 18.77\, L_1 
     - 40.035 + 114.4\, x 
  \nn \\ & & \; \mbox{}
     - 24.86\, x^2 - 53.39\, x^3
     + L_0 L_1\, ( 8.523\, L_0 + 269.4\, L_1 ) 
     - 26.63\, x L_0 
  \nn \\ & & \; \mbox{}
     + 270.0\, (1-x) L_1^2 
     - 21.55\, L_0 - 10.992\, L_0^2 - 32/27\: L_0^3 \, \big\}
\eea
and
\bea
  \lefteqn{
  \d_{\rm s}^{-1} P^{(2)}_{\rm g \g}(x)\! \:\: \cong \:\:
     (1-x) \: \big\{ \,
       32/27\: L_1^3 - 79.13\, L_1^2 + 87.22\, L_1
     + 1738 - 1580\, x }
  \nn \\[-0.2mm] & & \; \mbox{}
     - 160.0\, x^2 - 566.7\, x^3
     - L_0 L_1\, ( 549.5 + 1230\, L_0 + 433.2\, L_1 )
     + 2176\, L_0 
  \nn \\ & & \; \mbox{} 
     + 1123.7\, L_0^2 + 2400/27\: L_0^3 + 448/27\: L_0^4
     - 73.1409\, x^{\,-1} + 128/3\: x^{\,-1} L_0 \big\}
  \nn \\ & & \!\! \mbox{} + \Nf\, (1-x) \: \big\{ 
     - 32/9\: L_1^2 + 16.38\, L_1 + 68.10 - 36.42\, x + 56.95\, x^2
  \nn \\ & & \; \mbox{}
     - 44.10\, x^3
     - L_0 L_1\, ( 16.18 + 38.33\, L_0 + 9.133\, L_1 )
     - 10.76\, L_0
  \nn \\ & & \; \mbox{}
     + 26.41\, L_0^2 - 64/27\: L_0^3 
     - 40.5597\, x^{\,-1} \big\} 
\eea
with $L_0 \equiv \ln x$ and $L_1 \equiv \ln (1-x)$. 
These parametrizations deviate from the lengthy full expressions by about 
0.1\% or less, an accuracy which should be amply sufficient for practical 
applications. On the other hand, the exact expression is very compact for the 
NNLO pure-singlet splitting function, 
\bea
  \lefteqn{
  \d_{\rm s}^{-1} P^{(2)}_{\rm ps \g}(x)\! \:\: = \:\: 
     4/3\: \Nf\, \big\{ \, 2464/81\: x^{-1}  -  432 - 72\, x 
     + 38360/81\: x^2 } \nn \\
  & & \mbox{} - L_0 \, ( 344 + 368\, x + 3584/27\: x^2 ) 
      -  L_0^2 \, ( 144 + 104\, x + 224/9\:  x^2 ) \nn \\
  & & \mbox{} - L_0^3 \, ( 16 - 16\, x - 128/9\: x^2 ) 
     -  L_0^4 \; 8/3 \: (1 - 2\, x) \big\} \:\: .
\eea
Recall that $P_{\rm q\g}$ in Eqs.~(\ref{Pmat}) and (\ref{Pgexp}) is given by 
$P_{\rm q\g} = \d_{\rm s} / \d_{\rm ns}\, P_{\rm ns \g} + P_{\rm ps \g}$ 
with $\d_{\rm s} \,=\, 3 \,\sum_{j=1}^{\Nf}\, e_{q_j}^{\,2}$.
A pure-singlet term enters at order $\a \a_{\rm s}^2$ for the 
first time.

\begin{figure}[bht]
\centerline{\epsfig{file=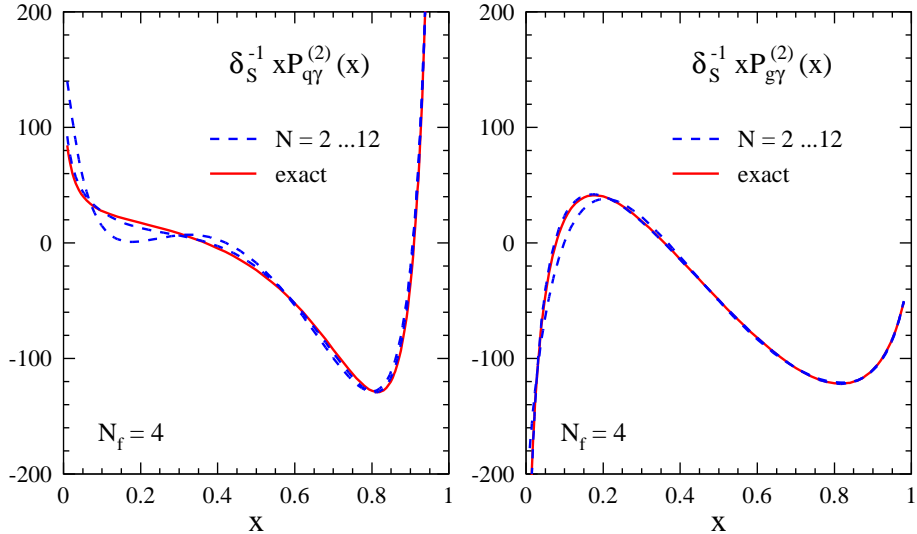,width=12.3cm,angle=0}}
\caption{\label{fig1}
The exact results for the $a_{\rm em} a_{\rm s}^2$ photon-quark (left) and 
photon-gluon (right) splitting functions (multiplied by $x$) in the \MSb\ 
scheme, compared with the (dashed) estimated error bands based on the lowest 
six even-integer moments \cite{Moch:2001im}.}
\end{figure}

\begin{figure}[thb]
\vspace*{-1mm}
\centerline{\epsfig{file=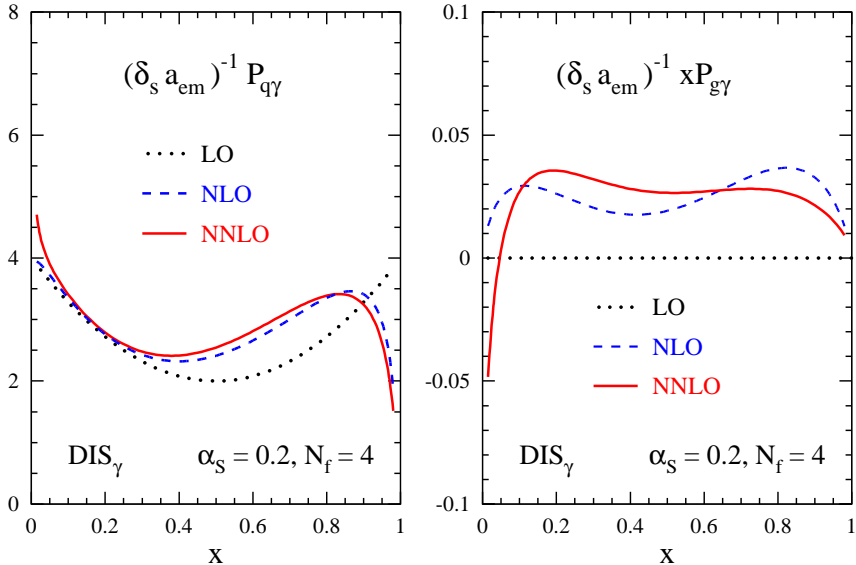,width=12cm,angle=0}}
\vspace*{-1mm}
\caption{\label{fig2}
The perturbative expansion (\ref{Pgexp}) of the photon-quark (left) and 
photon-gluon (right, multiplied by $x$) splitting functions in the DIS$_{\g}$ 
factorization scheme for typical values of $\a_{\rm s}$ and $\Nf$. 
Note the very different scales of the two graphs.
} 
\end{figure}

\vspace*{-7mm}
Our complete results for $P_{\rm p\g}^{(2)}$, p = q, g, are compared in Fig.~%
\ref{fig1} with the previous approximations of Ref.~\cite{Moch:2001im}. 
In Fig.~\ref{fig2} these results are combined, after transformation to the 
DIS$_{\g}$ scheme \cite{GRVg1,Moch:2001im}, with the lower-order splitting 
functions. 
As indicated by the previous approximate results, the perturbative expansion is
well-behaved at least at $x \gsim 0.1$. For a further discussion, including the
small-$x$ limit, the reader is referred to Ref.~\cite{MVVprep}.   
%
%

\end{document}